\documentclass{PoS}
\usepackage{amsmath, amsfonts,amssymb}

\newcommand{\be}{\begin{equation}}
\newcommand{\ee}{\end{equation}}
\newcommand{\benn}{\nonumber\begin{equation}}
\newcommand{\eenn}{\nonumber\end{equation}}
\def\bea{\begin{eqnarray}} \def\eea{\end{eqnarray}}
\def\beann{\begin{eqnarray*}} \def\eeann{\end{eqnarray*}}

\def\lsim{\raise0.3ex\hbox{$<$\kern-0.75em\raise-1.1ex\hbox{$\sim$}}}
\def\gsim{\raise0.3ex\hbox{$>$\kern-0.75em\raise-1.1ex\hbox{$\sim$}}}
\def\slash#1{#1\!\!\!\!\!/\!\,\,}
\def\Dslash{\slash D}

\title{\vspace*{-1cm}
\begin{flushright}
\texttt{\footnotesize CERN-PH-TH/2011-020\\ HIP-2011-05/TH}
\end{flushright}
\vfill
Numerical properties of staggered overlap fermions
}

\ShortTitle{Staggered overlap fermions}

\author{\speaker{Philippe de Forcrand} \\ 
Institute for Theoretical Physics, ETH Z\"urich, CH-8093 Z\"urich, Switzerland \\
and \\
CERN, Physics Department, TH Unit, CH-1211 Geneva 23, Switzerland \\
        E-mail: \email{forcrand@phys.ethz.ch}}

\author{Aleksi Kurkela\\
Institute for Theoretical Physics, ETH Z\"urich, CH-8093 Z\"urich, Switzerland \\
and\\
Department of Physics, McGill University, 3600 rue University, Montr\'eal, QC H3A 2T8, Canada\\
        E-mail: \email{aleksi.kurkela@mcgill.ca}}
\author{Marco Panero\thanks{Currently supported by the Academy of Finland, project 1134018.}\\
Institute for Theoretical Physics, ETH Z\"urich, CH-8093 Z\"urich, Switzerland \\
and\\
Department of Physics and Helsinki Institute of Physics, University of Helsinki,\\ FIN-00014 Helsinki, Finland\\
        E-mail: \email{marco.panero@helsinki.fi}}

\abstract{
We report the results of a numerical study of staggered overlap fermions, following
the construction of Adams which reduces the number of tastes from 4 to 2 without fine-tuning.
We study the sensitivity of the operator to the topology of the gauge field, its locality and
its robustness to fluctuations of the gauge field. We make a first estimate of the computing
cost of a quark propagator calculation, and compare with Neuberger's overlap.
}

\FullConference{The XXVIII International Symposium on Lattice Field Theory, Lattice2010\\
		June 14-19, 2010\\
		Villasimius, Italy}

\begin{document}

\section{Introduction}

The staggered Dirac operator,
\be
\Dslash_s = \frac{1}{2} \sum_\mu \eta_\mu (V_\mu - V_\mu^\dagger) \;\;\; \mbox{with:}\;\; (V_\mu)_{xy} = U_\mu(x) \delta_{y,x+\hat\mu} \;\;\; \mbox{and}\;\; \eta_\mu(x) = (-1)^{\sum_{\nu<\mu} x_\nu},
\ee
is the most computationally efficient way to discretize the Dirac operator, 
and is commonly used in large-scale lattice QCD simulations.
Yet, it leads to 4 degenerate quark ``tastes'' in the continuum limit, and the determinant is
raised to the power $N_f/4$ in $N_f$-flavor simulations. The systematic error associated with 
this ``rooting'' is a subject of hot debate. 
In the literature, two different types of staggered-like operators which avoid rooting by representing only two tastes have been proposed:
\\
$i)$ minimally doubled staggered fermions~\cite{Karsten-Wilczek,Creutz-Borici}, which represent 
2 tastes with minimal fine-tuning; \\
$ii)$ staggered overlap fermions~\cite{Adams1,Adams2}, where the degeneracy of the spectrum 
is lifted by a taste-dependent mass term, and the resulting operator
is used as the kernel in Neuberger's overlap~\cite{Neuberger}.

Here we study the numerical properties of $(ii)$. In spite of the additional complexity of the overlap, it may be simpler than the multiple fine-tuning required in $(i)$~\cite{Capitani}. Note also that the overlap kernel entering $(ii)$ can also be used without overlap, at the expense of fine-tuning. 
We first motivate the overlap kernel devised in \cite{Adams1} and study its topological properties, then
consider the staggered overlap operator of \cite{Adams2}.

\section{Index of overlap kernel}

\begin{figure}
\centerline{
\includegraphics[width=0.40\textwidth]{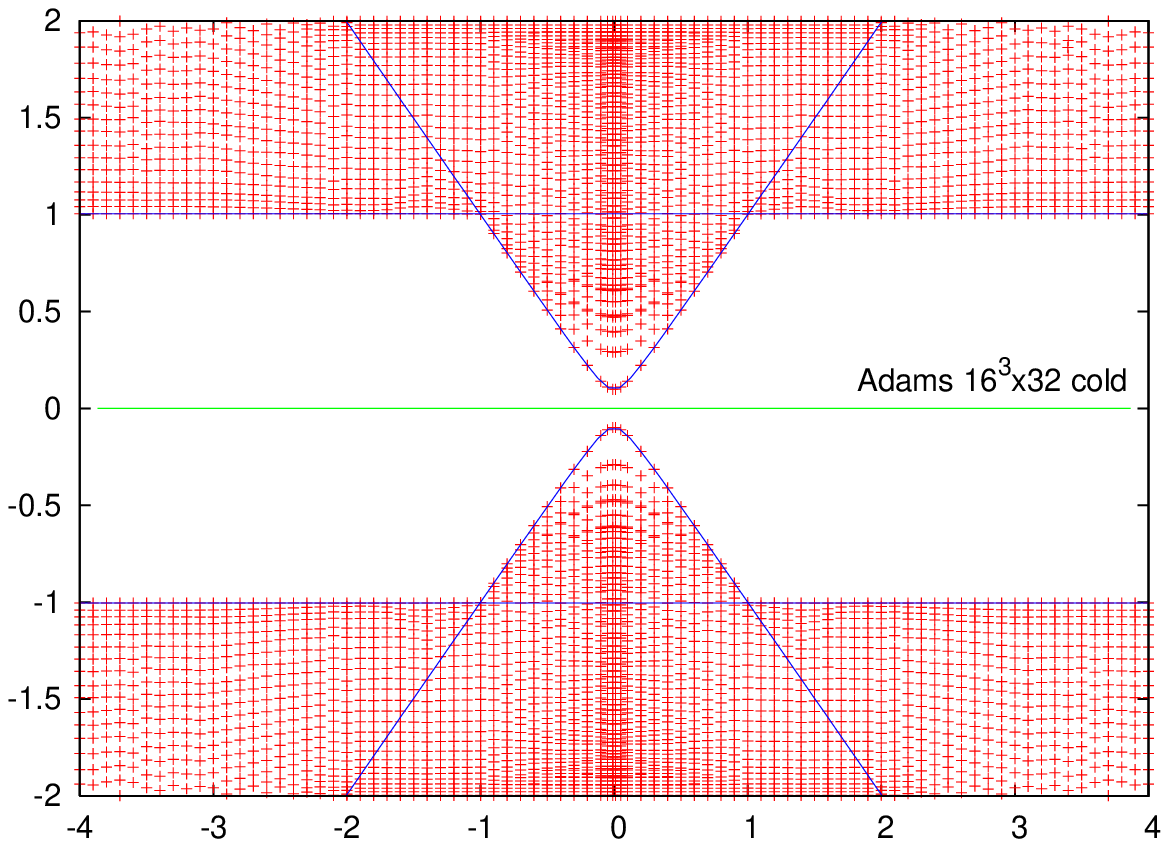}
\hspace*{1.5cm}
\includegraphics[width=0.40\textwidth]{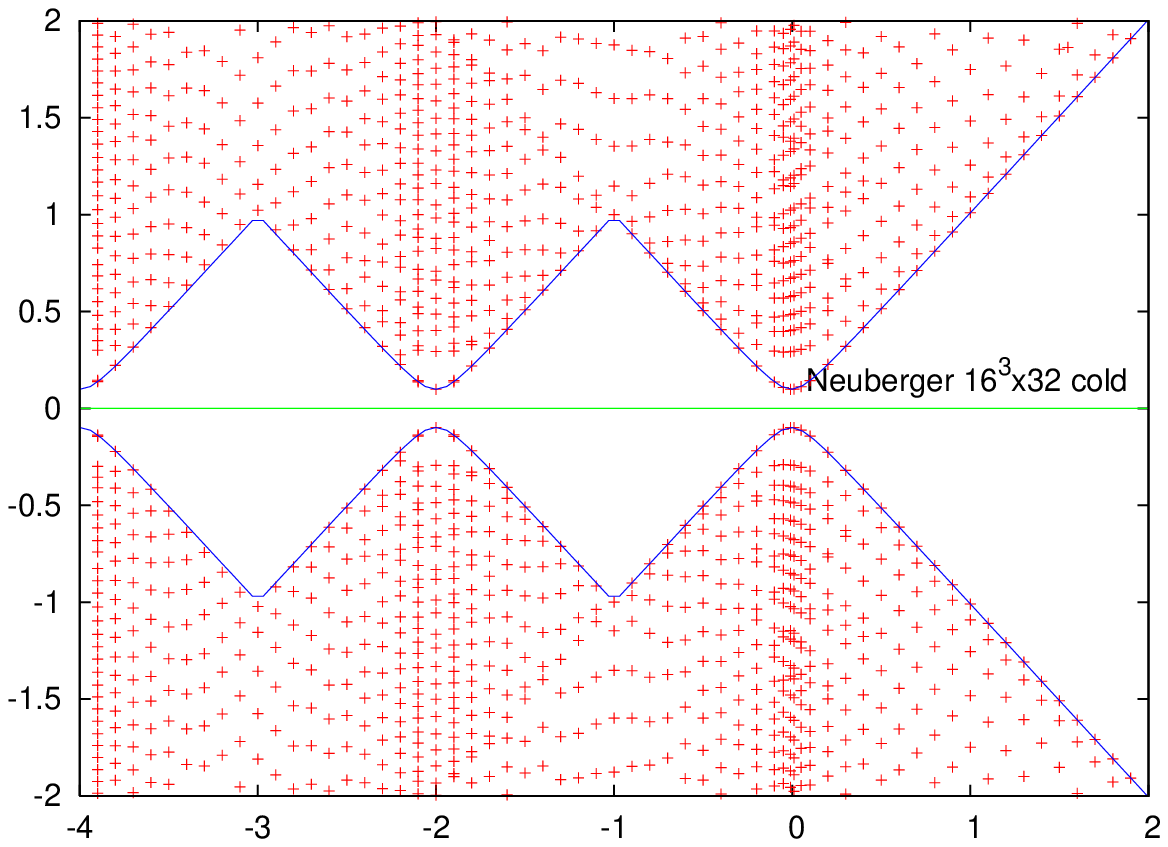}
}
\centerline{
\includegraphics[width=0.40\textwidth]{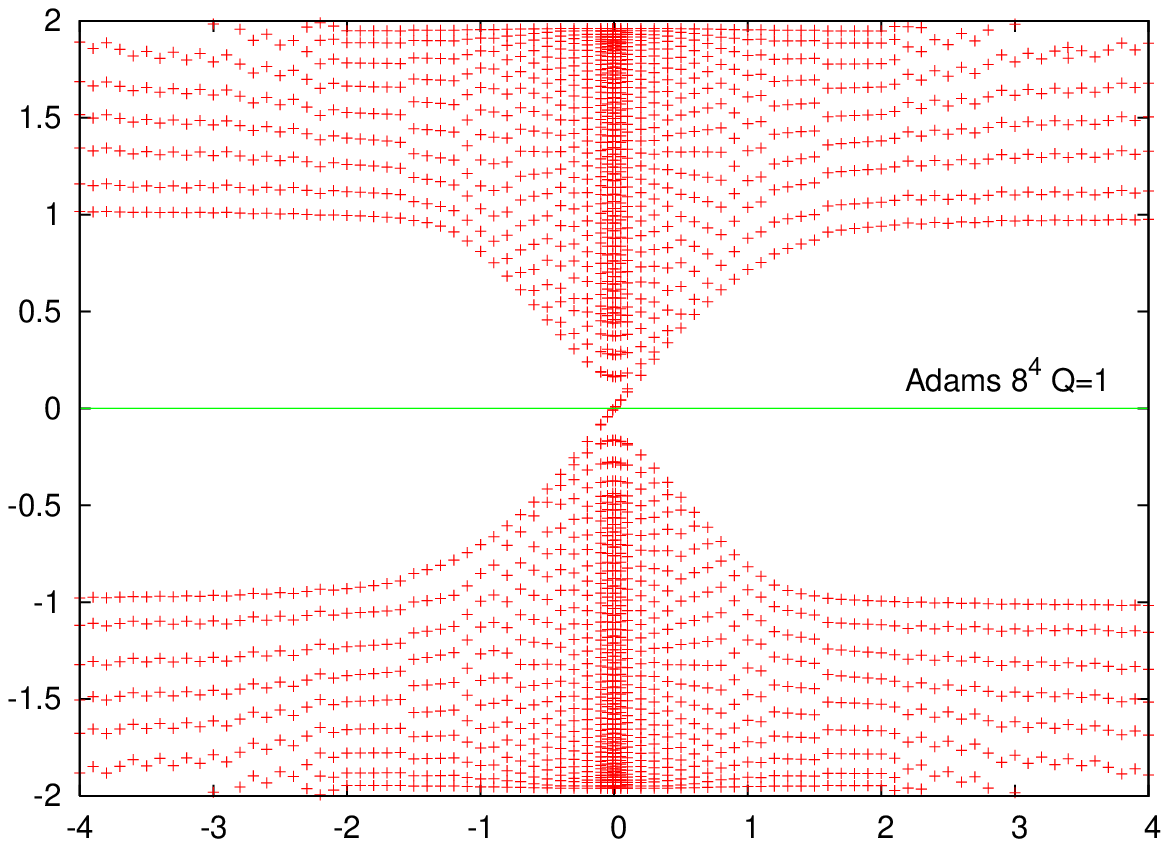}
\hspace*{1.5cm}
\includegraphics[width=0.40\textwidth]{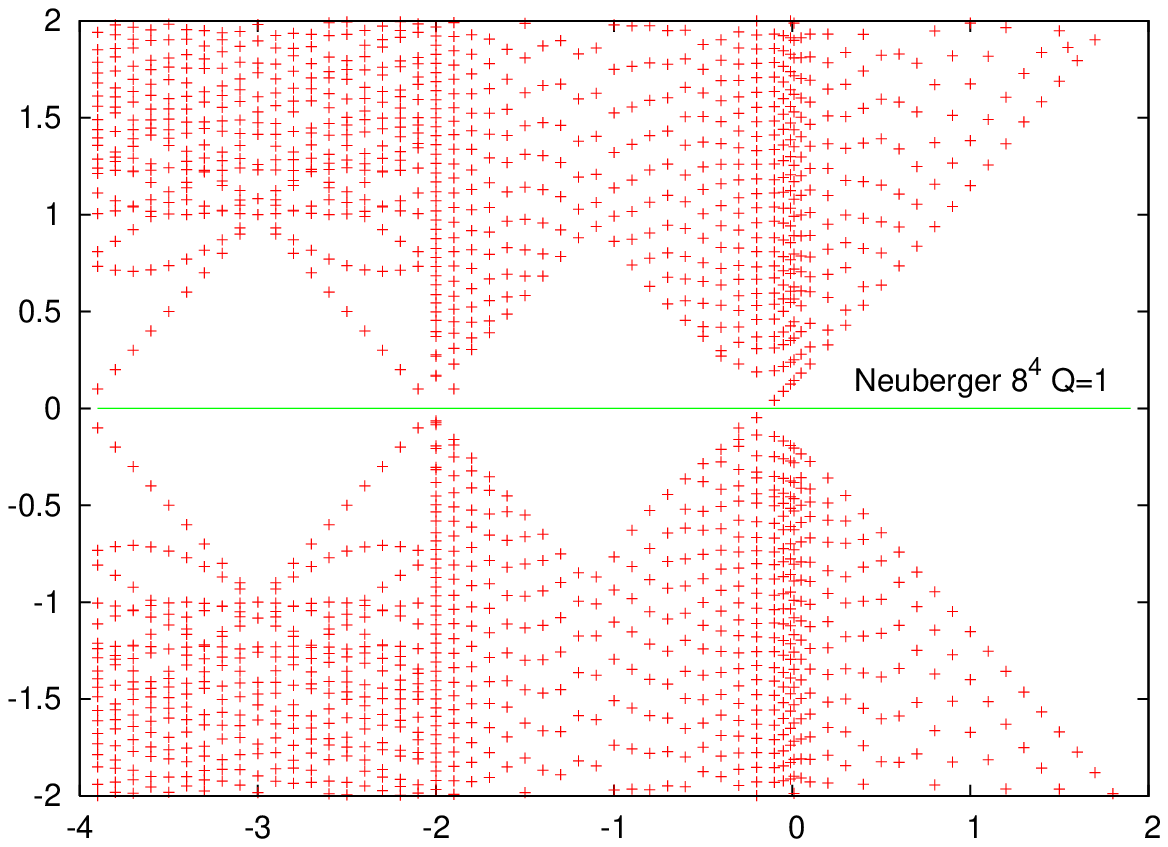}
}
\centerline{
\includegraphics[width=0.40\textwidth]{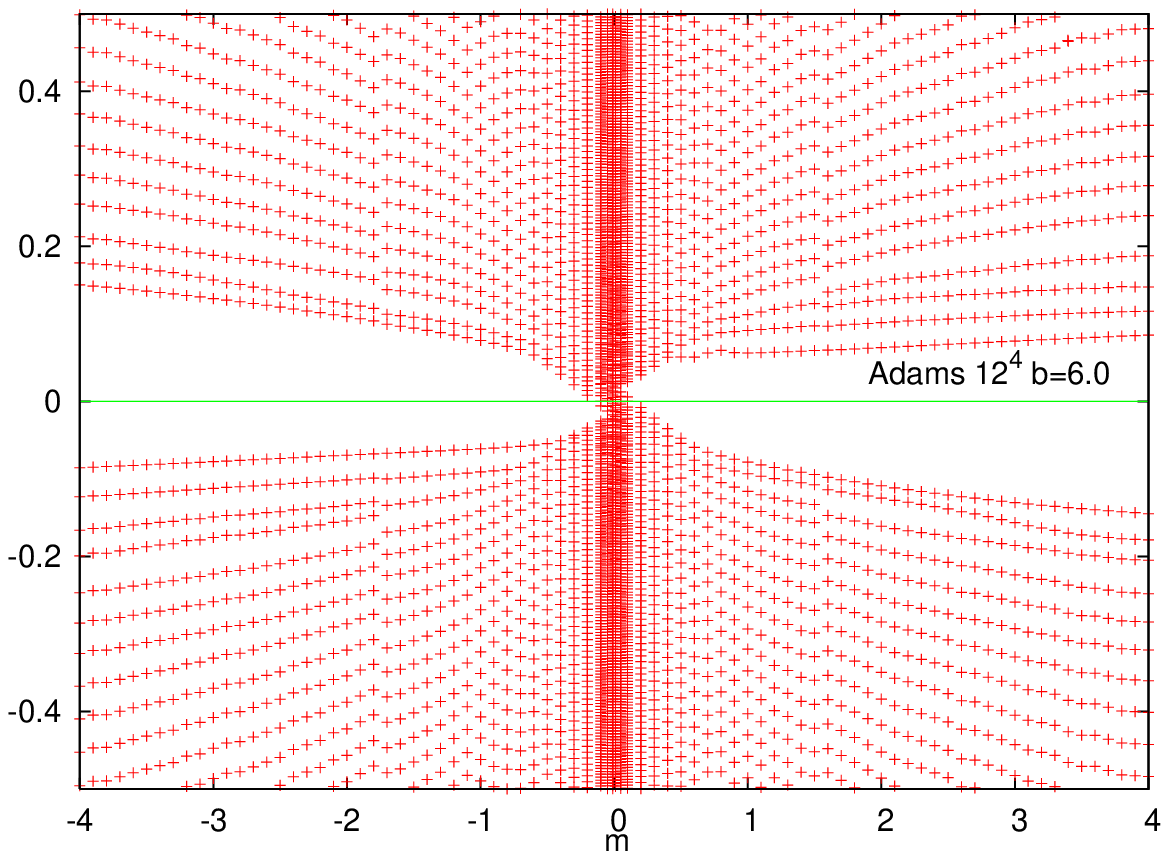}
\hspace*{1.5cm}
\includegraphics[width=0.40\textwidth]{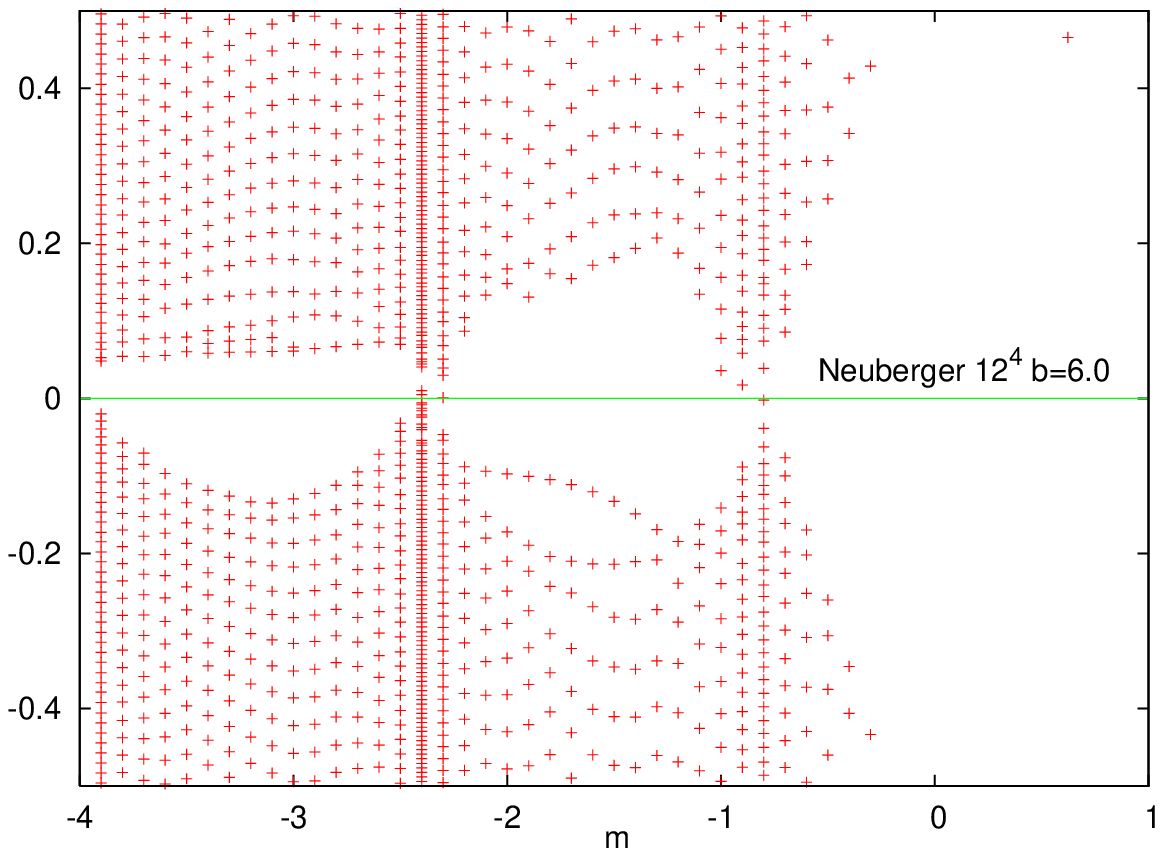}
}
\caption{Flow of eigenvalues {\emph vs.} $m$ for Adams' operator (left) and Neuberger's operator (right).}
\label{fig:flow}
\end{figure}

The index of a gauge field configuration is obtained from the flow of the eigenvalues of
$H(m) = \gamma_5 (\Dslash + m) $ 
as a function of $m$. 
To probe 
the topological properties of the gluon field, it is essential that
the varying term $m \gamma_5$ be a taste singlet. This is not the case for staggered fermions, if one makes for $\gamma_5$
the customary choice $\gamma_5 \to \epsilon(x,y) = (-1)^{\sum_\mu x_\mu} \delta_{x,y}$, which has the
decomposition $\epsilon = \gamma_5 \otimes \gamma_5$ in spin $\otimes$ taste space.
Instead, Adams proposes to trade $\epsilon$ for $\Gamma_5 = \eta_5 C$, where
$C = \frac{1}{24} \sum_{ijkl = {\rm perm}(1234)} C_i C_j C_k C_l$, $C_\mu = \frac{1}{2} (V_\mu + V_\mu^\dagger)$, is the symmetrized sum of 4-link
parallel transporters connecting a site to its opposite in an elementary hypercube, and $\eta_5(x,y) = \prod_{\mu=1}^{4} \eta_\mu = (-1)^{x_1+x_3} \delta_{x,y}$ is
the corresponding product of 4 phase factors. $\Gamma_5$ has the spin $\otimes$ taste 
decomposition $\gamma_5 \otimes {\bf 1}$: it is a taste singlet, which allows $m\Gamma_5$ to
probe the topology of the gauge field.

In Figs.~\ref{fig:flow}, we compare 
the flow of eigenvalues with $m$ for Adams' operator
\be
H_A(m) = \epsilon \Dslash_s + m \Gamma_5
\ee
and for Neuberger's operator 
$H_W(m) = \gamma_5 (\Dslash_W + m)$
(where $\Dslash_s$ and $\Dslash_W$ are the zero bare mass staggered and Wilson Dirac operators, respectively), on the same $SU(3)$ gauge field configurations.

The top pair of figures corresponds to the free case (the solid lines are analytic results).
In the second set of figures, the gauge field is that of a smooth, cooled instanton. The eigenvalue
flow shows one crossing in Neuberger's case, two nearly degenerate crossings in Adams' case,
reflecting the corresponding number of flavors.
The third pair of figures corresponds to a thermalized configuration ($\beta=6$ with Wilson gauge action).
While the pattern of eigenvalue flow is less clear for both operators, it is remarkable that the eigenvalue gap in Adams' case seems to persist for arbitrarily large values of $|m|$~\footnote{
In Adams' case, eigenvalues come in pairs $\lambda(m) \leftrightarrow -\lambda(-m)$, because
$\epsilon H_A(m) \epsilon = -H_A(-m)$.}, 
while in Neuberger's case additional eigenvalue crossings appear at negative values of $m$ (corresponding to theories with a larger number of flavors), shifted from their free-field values by gauge field fluctuations.

\section{Adams' staggered overlap operator}

\begin{figure}
\centerline{
\includegraphics[width=0.75\textwidth,clip=true]{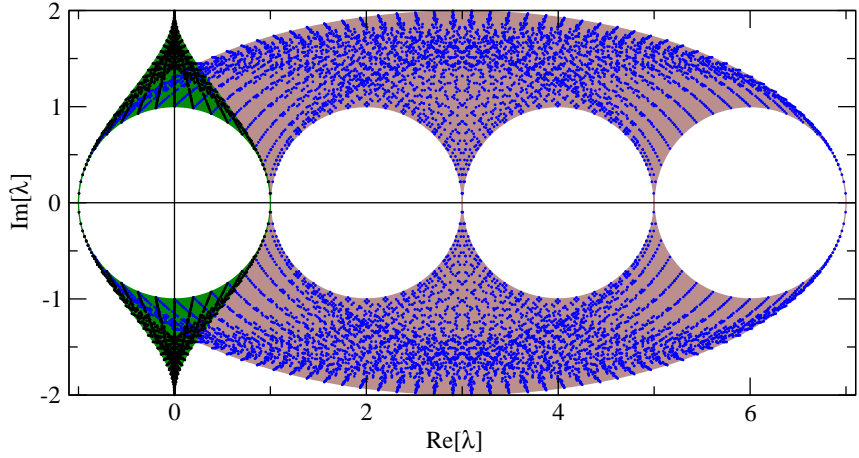}
}
\caption{Spectrum of the free Wilson operator (blue) and of the free Adams' operator (green).}
\label{fig:free_spectra}
\end{figure}

Having constructed a Hermitian kernel $H_A(m) = \epsilon \Dslash_s + m \Gamma_5$ sensitive to the topology of the gauge field, one can plug this kernel into Neuberger's overlap: $\Dslash_{ov} = {\bf 1} + \gamma_5 ~{\rm sign}(H(-m_0)) = {\bf 1} + D/\sqrt{D^\dagger D}$, 
where $D = \gamma_5 H$ and $D^\dagger = \gamma_5 D \gamma_5$. 
Thus, Adams' staggered overlap operator is
\be
\Dslash_{sov} \equiv {\bf 1} + \epsilon ~{\rm sign}(H_A(-m_0)) = {\bf 1} + D_A/\sqrt{D_A^\dagger D_A},
\ee
with $D_A = \epsilon H_A(-m_0) = \Dslash_s - m_0 \epsilon \Gamma_5$.
Note that the mass term $-m_0 \epsilon \Gamma_5$ has the spin $\otimes$ taste decomposition ${\bf 1} \otimes \gamma_5$: it is ``flavored''. If one considers low-momentum eigenstates $\tilde\Psi$ of $\Dslash_s$, satisfying $\Dslash_s \tilde\Psi \approx 0$, they will obey
$\langle \tilde\Psi^\dagger (\epsilon \Gamma_5) \tilde\Psi \rangle \approx \pm 1$ depending on their
taste content. Of the initial 4 tastes, two combinations will give $+1$ and become physical, light modes of
the overlap operator; the other two combinations will give $-1$ and become heavy doublers.

This is 
clear in the free case: Fig.~\ref{fig:free_spectra} shows the spectrum of $D_A$ and of the Wilson operator $D_W$, for 
$m_0=1$ and a free field. The splitting of the 4 tastes into 2 pairs is achieved in a symmetric way, more elegant than the reduction from 16 flavors to 1 in the Wilson case.
Moreover, 
since Adams' kernel has a spectrum
already much closer to the unit circle than Wilson's operator,
one may expect a smaller number of operations to achieve the unitary projection $D/\sqrt{D^\dagger D}$ of the kernel operator $D$ in Adams' case than in Neuberger's case. This is investigated in Sec.~\ref{sec:cost}.

\section{Locality}

\begin{figure}
\centerline{
\includegraphics[width=0.50\textwidth]{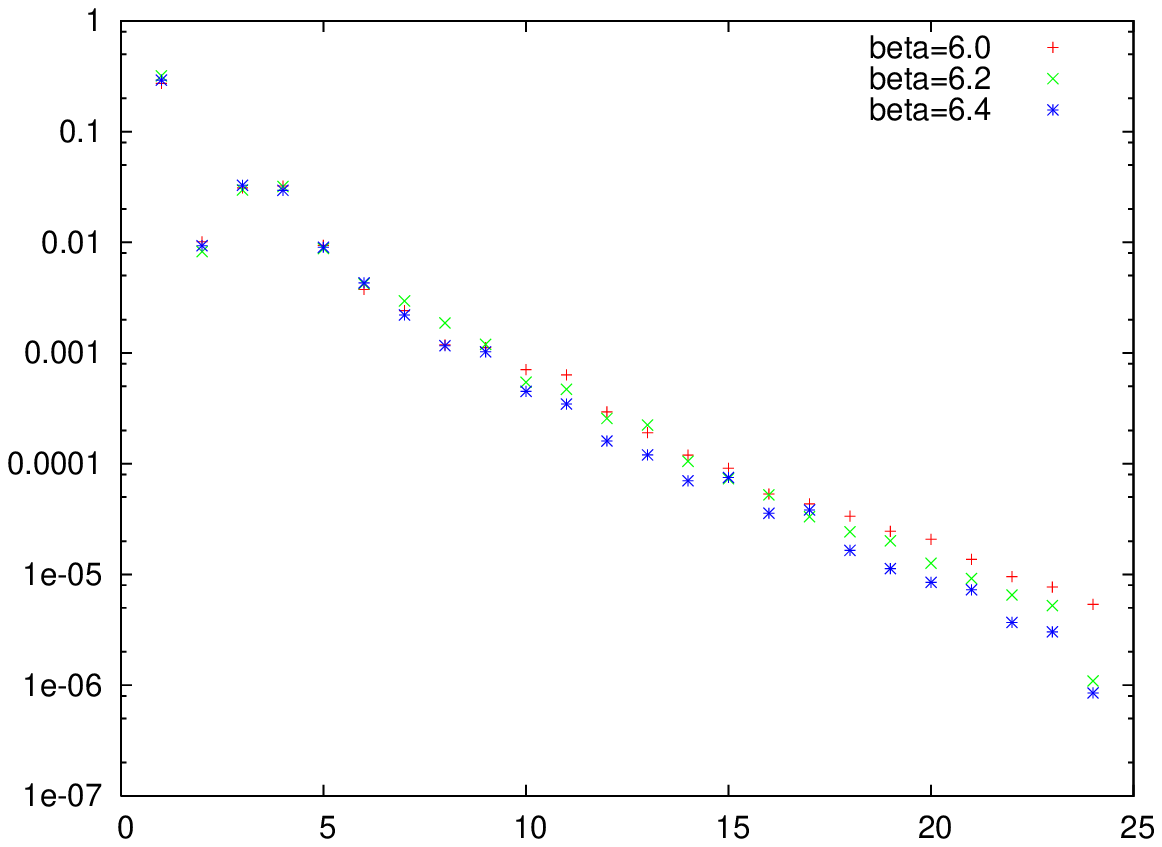}
\includegraphics[width=0.50\textwidth]{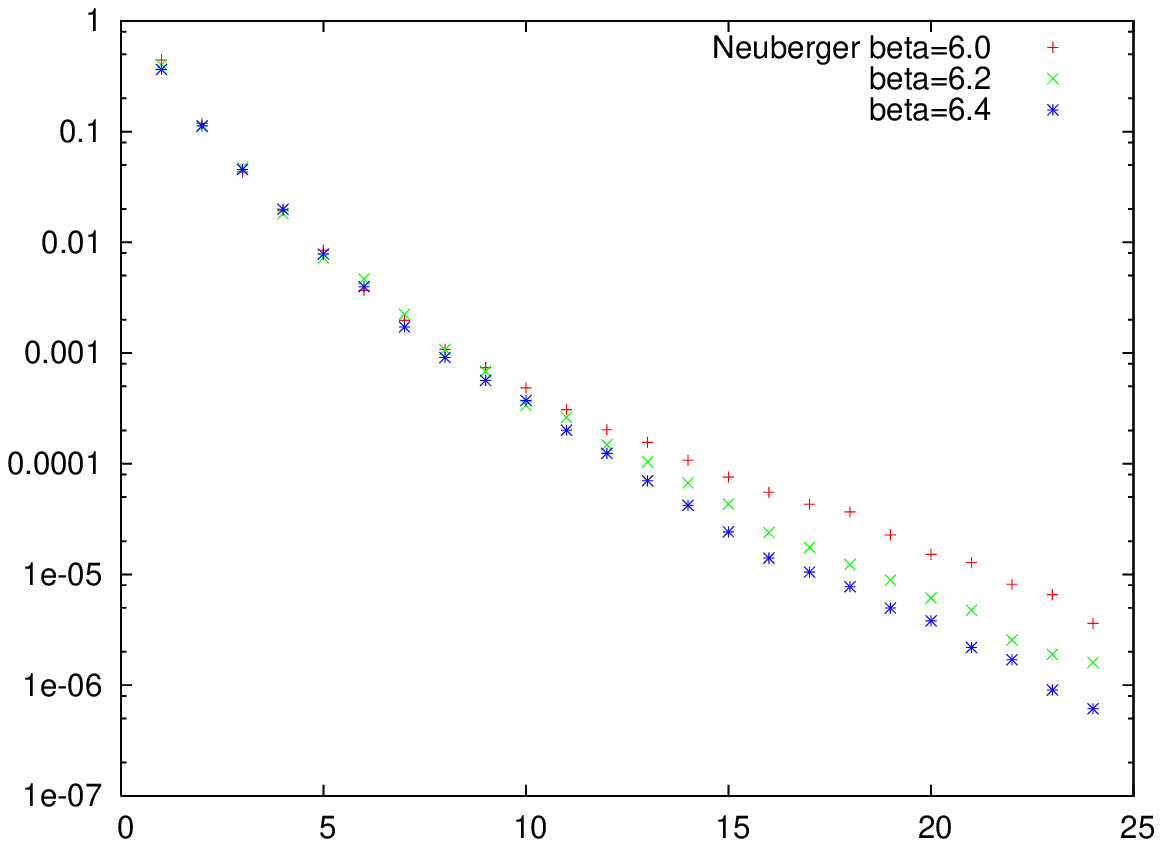}
}
\caption{Maximum magnitude ${\rm max}_y |D_{ov}(x,y)|$ of the Dirac operator matrix elements versus Manhattan distance $|x-y|$, for Adams' operator (left) and for Wilson's operator (right).}
\label{fig:locality}
\end{figure}
First, we compare the locality of Adams' overlap operator with that of Neuberger's. In both cases, the
matrix elements $D_{ov}(x,y)$ are non-zero for arbitrarily distant sites $x$ and $y$, 
as the overlap operator is not ultra-local. What matters, however, is the decrease in magnitude
of $|D_{ov}(x,y)|$ with the distance $|x-y|$, which should be bounded by $\exp(-|x-y|/(c a))$, where $(c a)$
is a localization 
length proportional to the lattice spacing $a$, and thus shrinking to 
zero in the continuum limit.

Fig.~\ref{fig:locality} shows the maximum magnitude $\max_y |D_{ov}(x_0,y)|$ versus the Manhattan distance $|x_0-y|$, chosen to follow the conventions of Ref.~\cite{locality} for the Neuberger operator. The left 
figure corresponds to Adams' overlap operator, the right one to Neuberger's, on the same gauge configurations at 3 values of $\beta$. While Adams' operator behaves differently at short distance because of the 4-link transporters, at large distances the decay of the matrix elements is
exponential as in Neuberger's case, with a similar localization length.
\footnote{This happens even though the
kernel of Adams' operator is much less local than that of Neuberger's: 
a less ultralocal kernel may lead to a more local overlap operator~\cite{Bietenholz_Durr}.}

\section{Robustness to gauge fluctuations and efficiency}
\label{sec:cost}

\begin{figure}
\centerline{
\includegraphics[width=0.39\textwidth]{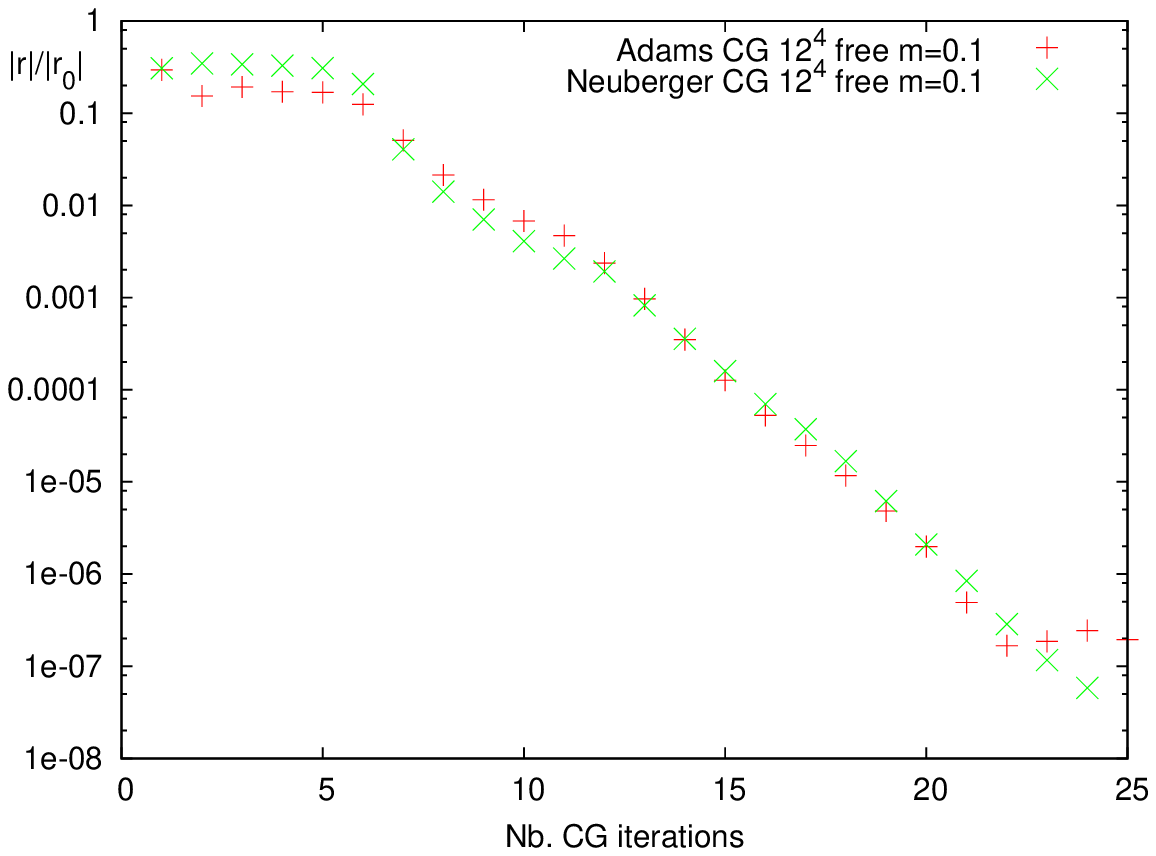}
\includegraphics[width=0.39\textwidth]{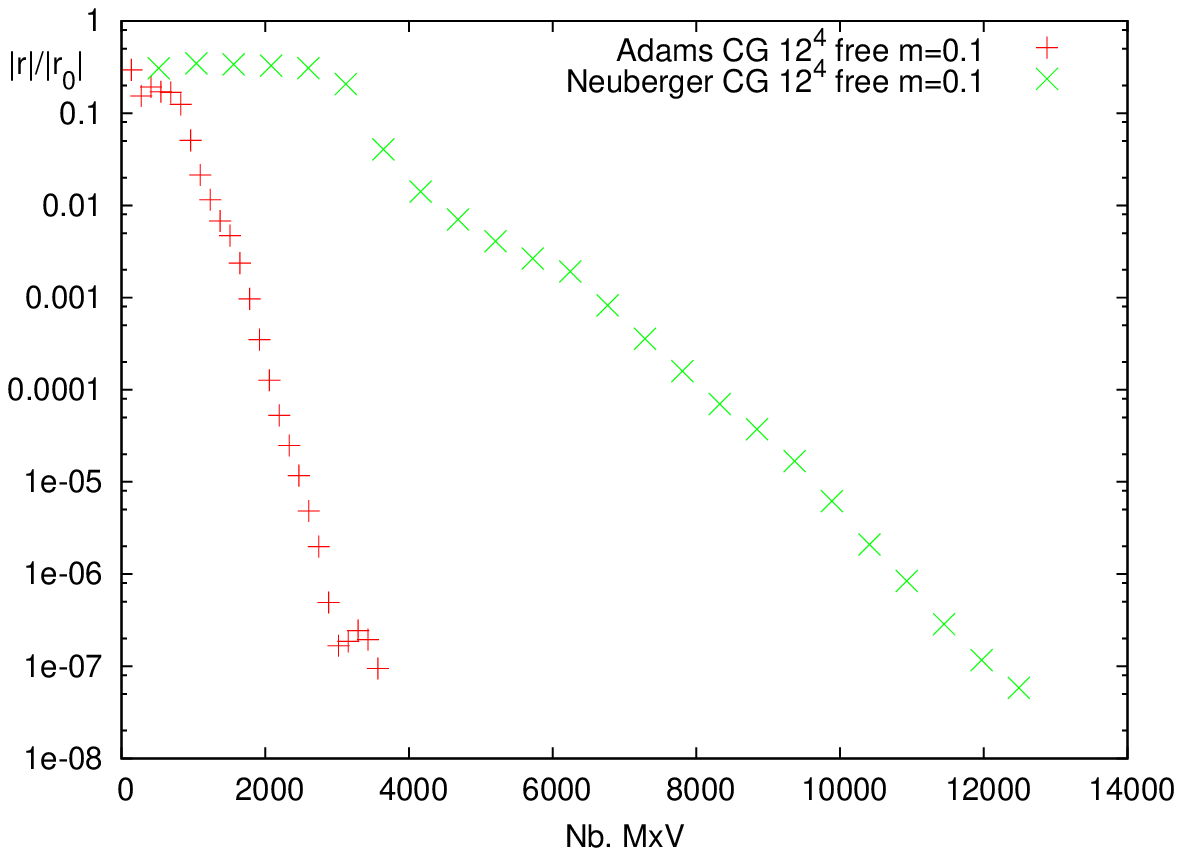}
\includegraphics[width=0.39\textwidth]{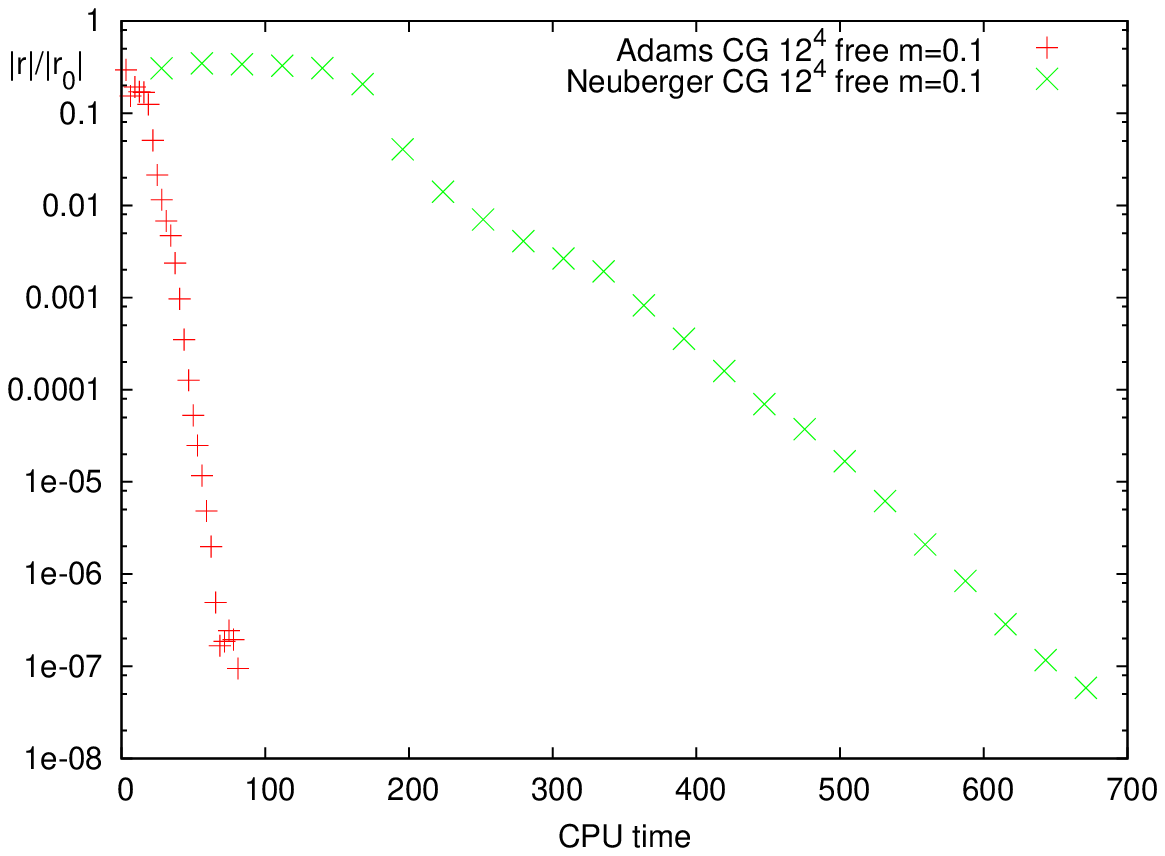}
}
\centerline{
\includegraphics[width=0.39\textwidth]{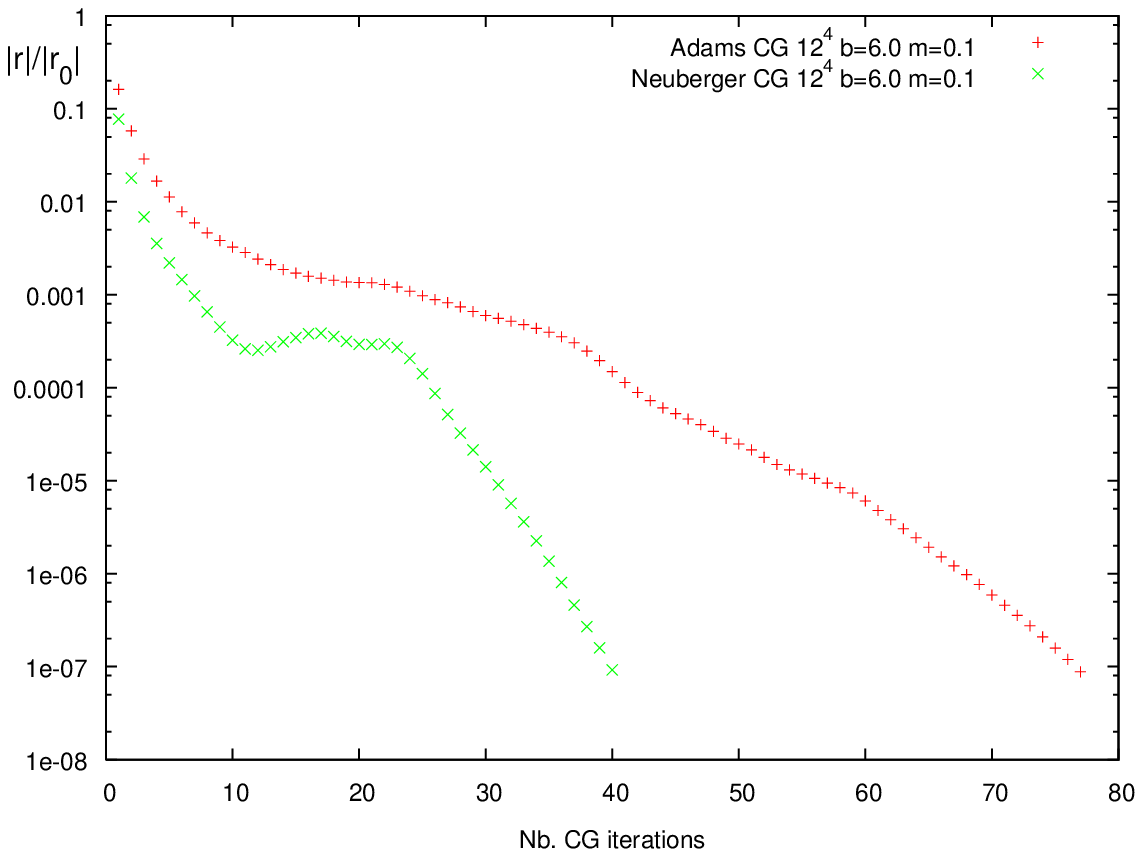}
\includegraphics[width=0.39\textwidth]{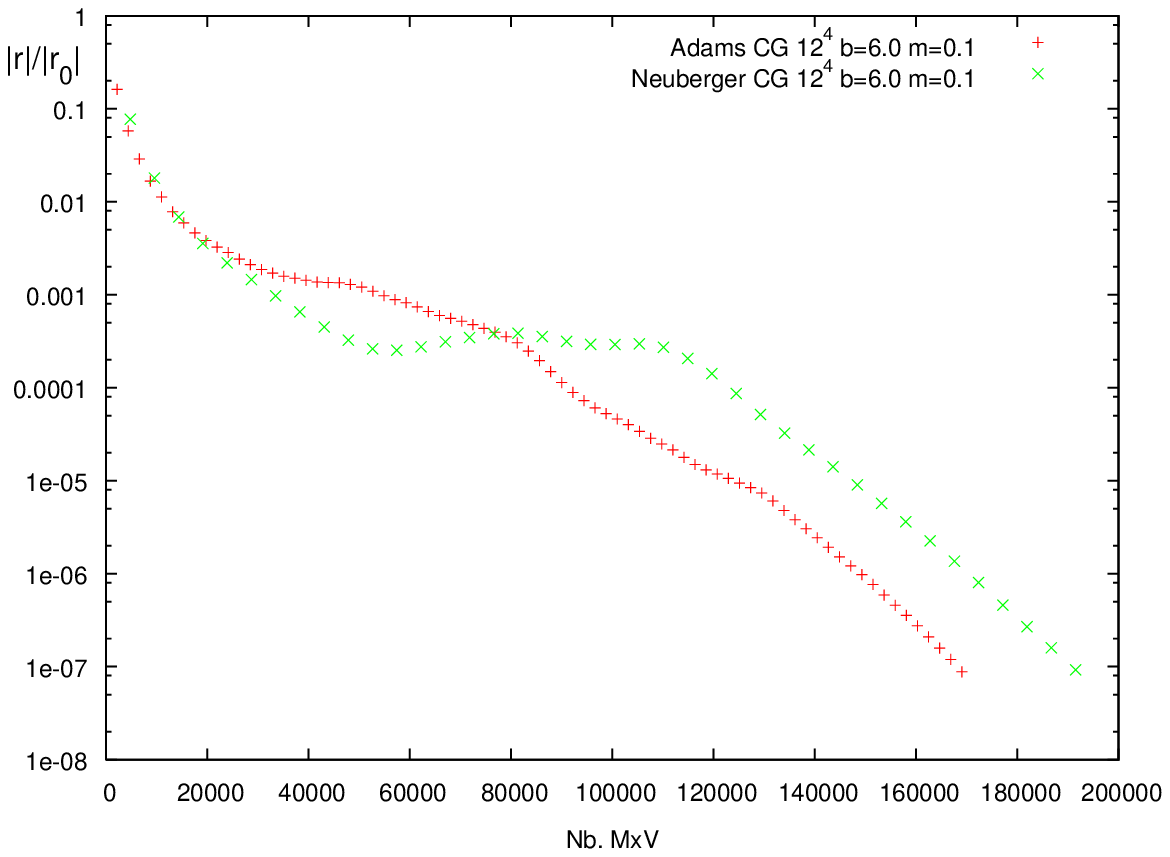}
\includegraphics[width=0.39\textwidth]{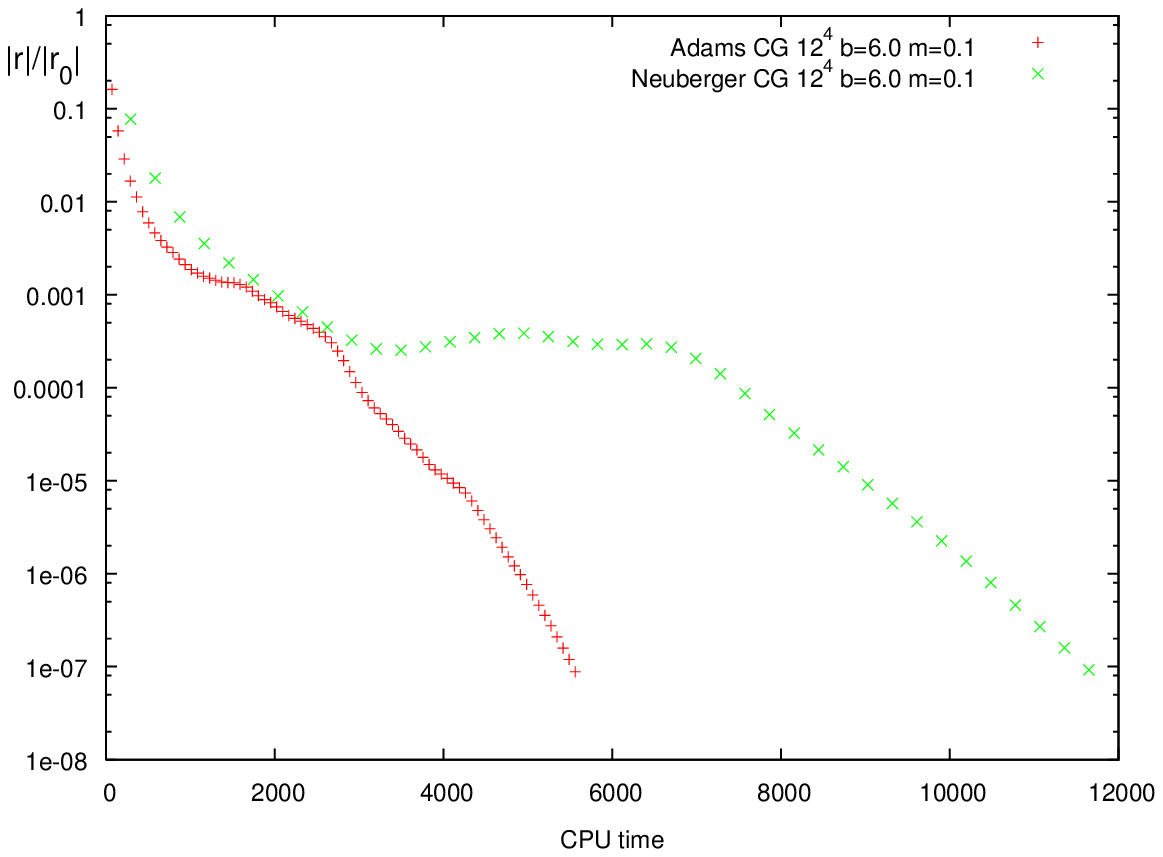}
}
\hspace*{0.6cm} Outer CG iter.
\hspace*{4.4cm} MxV
\hspace*{5.0cm} {\bf CPU}
\caption{Computer cost of one overlap propagator, measured in outer CG iterations (left), matrix-vector multiplications (middle) and CPU time (right). The values for Adams' operator are shown in red, those for Neuberger's operator in green. The gauge field is the free field (top), or a $\beta=6.0$ configuration (bottom).}
\label{fig:cost}
\end{figure}

We have studied the computational cost of a quark propagator calculation with Adams' operator and compared it to Neuberger's propagator (for one component) on the same gauge field background, 
and with the same numerical approach. 
As the matrix to invert is 4 times smaller, and its spectrum is closer to the unit circle, at least in the free case (Fig.~\ref{fig:free_spectra}),
Adams' operator may be computationally cheaper.

The propagator is obtained as the solution of
$(D_{ov}+m)^\dagger (D_{ov}+m) x = (D_{ov}+m)^\dagger b$, using a conjugate gradient iterative solver, using the following simple and robust method ~\cite{Borici}: at each iteration of this outer CG, 
${\rm sign}(H)$ is applied to a vector $v$ through a 
Lanczos process, building a tridiagonal matrix. Its eigenvalues are representative of those of $H$, and we replace them by their sign. The results are presented in Fig.~\ref{fig:cost}, for a 
free field (top) and a $\beta=6$ configuration (bottom). 
The 3 figures in each row show the relative norm of the residual,
$|r|/|r_0|$, 
\emph{vs.} outer CG iterations (left), number of matrix-vector multiplications (middle) and CPU time (right).

In the free field case, the CPU time to 
find the solution for Adams' operator is almost an order of magnitude smaller than for Neuberger's, 
thanks to a construction of the sign function requiring fewer matrix-vector multiplications,
each with a smaller CPU cost.
The inversion of the unitary operator converges  
at the same rate, reflecting 
the similar spectral properties in the infrared.

The situation changes on a $\beta=6$ configuration. The outer CG now converges noticeably faster in Neuberger's case (left). This advantage is offset by the cost of the sign function, which still is cheaper in Adams' case (middle). Finally, the CPU time per matrix-vector multiplication is a factor ${\cal O}(2)$
smaller in Adams' case (Adam's matrix is one quarter the size of Neuberger's, with each site connected
to (8+16) neighbours, instead of 8 with 2 Dirac components). In total, the CPU time to find the solution is only
${\cal O}(2)$ times smaller in Adams' case.

\begin{figure}
\centerline{
\includegraphics[width=0.275\textwidth,clip=true]{./StaggeredBatman.eps}
\includegraphics[width=0.283\textwidth]{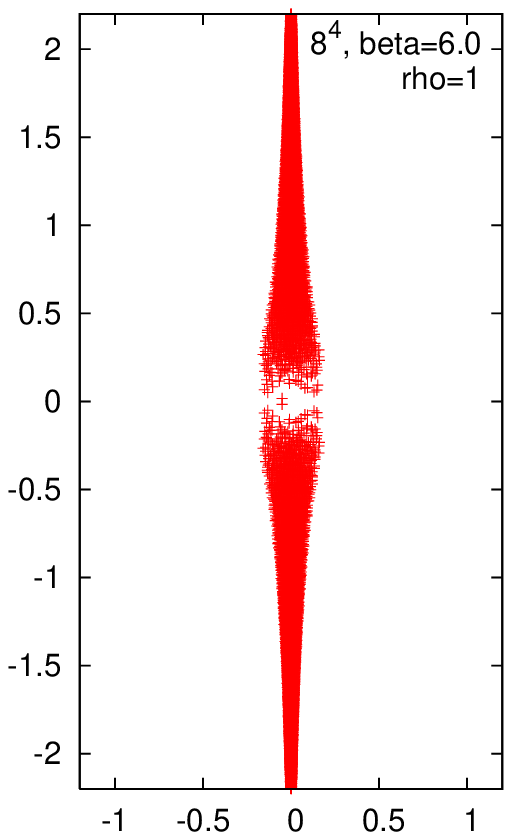}
\includegraphics[width=0.283\textwidth,clip=true]{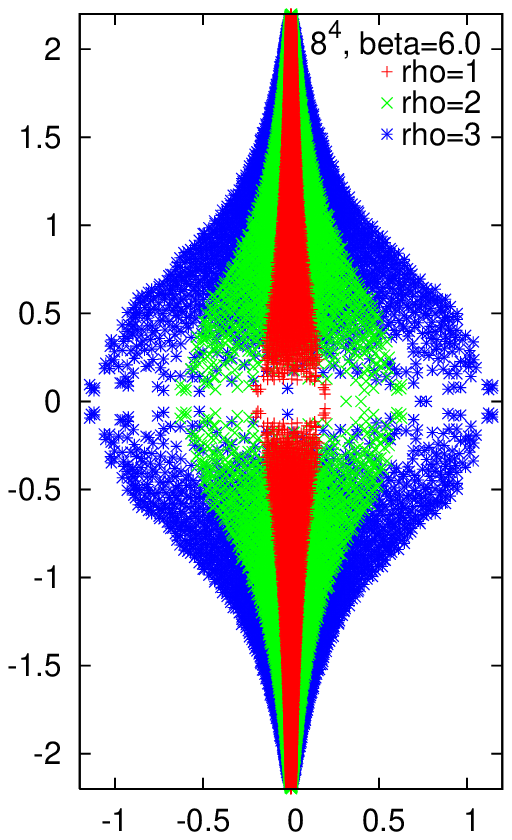}
\includegraphics[width=0.283\textwidth,clip=true]{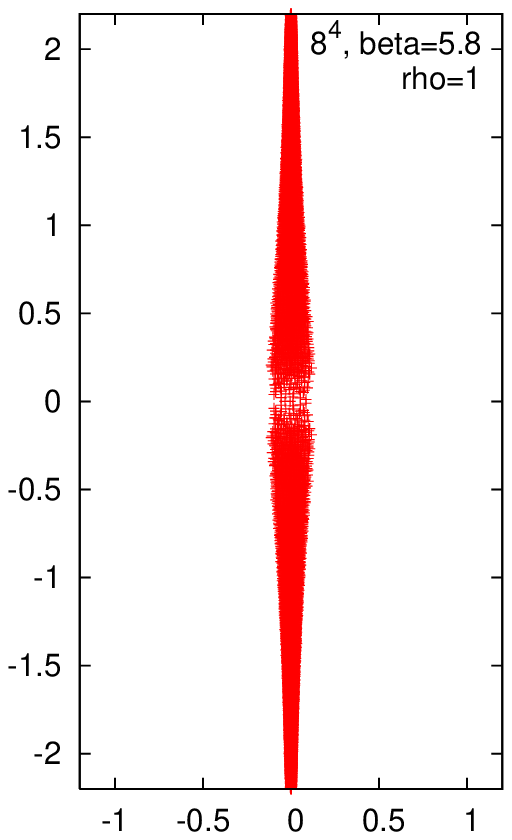}
}
\caption{Spectra of Adams' operator. Left to right: free field, $\beta=6.0$, $\beta=6.0$ with 
larger $\rho$, $\beta=5.8$.}
\label{fig:spectra}
\end{figure}

This loss of efficiency can be traced to changes in the spectrum of Adams' kernel $D_A$ in the presence
of gauge field fluctuations, as illustrated in Fig.~\ref{fig:spectra}. While the free spectrum (left)
is remarkably close to the unit circle, it is quite different at $\beta=6$ (2nd panel). The splitting
of the 4 tastes into 2 pairs is markedly reduced. The reason is that the taste-dependence of the mass
operator $\epsilon\Gamma_5$ is achieved via 4-link transporters: fluctuations in the gauge links are raised to the 4th power. It is the same reason for which the chirality $\langle \Psi^\dagger \Gamma_5 \Psi \rangle$ of near-zero modes of the ordinary staggered operator is so small~\cite{Hetrick}.
Here, one may attempt to restore the mass splitting of the pairs of tastes, by increasing the mass parameter $m_0$, called $\rho$ in Adams' Ref.~\cite{Adams2}. The effect of such increase is shown Fig.~\ref{fig:spectra} (3rd panel). The gap in the spectrum, which was the complete unit disk in the free case, and which shrank to a small but disk-like shape at $\beta=6$ for $m_0=1$, now becomes a very narrow
band. Unitary projection of the operator becomes more difficult, and after unitarization many modes are
present near the origin, which makes inversion more difficult as well. 

This figure also shows that the spectrum remains centered about the origin: changing $m_0$ is not the analogue of changing the mass in the Wilson operator, which shifts the whole spectrum. 
Rather, $m_0$ is the analogue of Wilson's hopping parameter $r$~\cite{Adams2}.

This is why the eigenvalue gap in the Hermitian operator $H_A(m)$ (Fig.~\ref{fig:flow}) persists for large values of $|m|$.
Shifting the whole spectrum of $D_A$ by 
a taste-independent mass term is also possible, but will destroy the symmetry of the spectrum about the origin
without any computational advantage. Finally, Fig.~\ref{fig:spectra} (right) shows how the gap in the spectrum of $D_A$ closes at $\beta=5.8$.

\section{Conclusion}

Our 
study shows that Adams' staggered overlap operator works as advertised: 
the taste-dependent mass operator in its kernel yields 2 massless tastes without fine-tuning, and the topology and locality properties are similar to Neuberger's operator.

On very smooth gauge configurations, the computer cost of a quark propagator is nearly an order of magnitude less than in Neuberger's case, 
but the 4-link transporters in the flavored mass term reduce this advantage to a factor ${\cal O}(2)$ on $\beta=6$ configurations. Another drawback of Adams' construction is that the continuous symmetry of the massless staggered
overlap operator is $U(1)$, not $SU(2)$ as one would wish for a 2-flavor chiral symmetry. 

The lack of robustness and of full chiral symmetry can both be addressed by modifying the mass operator, 
for example with 2-link transporters~\cite{Hoelbling} that reduce the number of light tastes to 1. Preliminary results~\cite{prepare} confirm our expectations, but do not bring the cost of staggered overlap fermions near that of ordinary staggered fermions: avoiding rooting still has its price.


\end{document}